# Ultrafast Spatial Hole Burning Dynamics in Monolayer $WS_2$: Insights from Time-resolved Photoluminescence Spectroscopy

Yichun Pan,[1] Liqing Zhu,[1] Yongsheng Hu,[2] Xin Kong,[1] Tao Wang,[1] Wei Xie,[2*] and Weihang Zhou[1*]

[1] Wuhan National High Magnetic Field Center and School of Physics, Huazhong University of Science & Technology, Wuhan 430074, China

[2] State Key Laboratory of Precision Spectroscopy, School of Physics and Electronic Science, East China Normal University, Shanghai 200241, China

To whom correspondence should be addressed: wxie@phy.ecnu.edu.cn and zhouweihang@hust.edu.cn

# Abstract

The transport of excitons lies at the heart of excitonic devices. Probing, understanding, and manipulating excitonic transport represents a critical step prior to their technological applications. In this work, we report experimental studies on the ultrafast nonlinear transport of excitons in monolayer $WS_2$. Under intense optical pumping, we observed an ultrafast spatial hole burning effect in the excitonic emission profile, followed by a re-brightening at even higher pumping density. By means of time- and spatially-resolved photoluminescence imaging spectroscopy, we revealed the underlying mechanism responsible for these nontrivial excitonic diffusion dynamics. Our results demonstrate that the combined effects of ultrafast exciton-exciton annihilation, efficient hole trapping by intrinsic sulfur vacancy defects, and laser-induced photo-oxidation govern the evolution of exciton transport under strong optical excitation. The observed dynamics are in excellent agreement with our diffusion model simulations, providing new insights into the nonlinear excitonic transport behaviors as well as their optical control mechanism in two-dimensional semiconductors.

# Introduction

As typical two-dimensional layered semiconductors, transition metal dichalcogenides (TMDCs) have received tremendous attention in recent years. Compared with bulk semiconductors, one of the most intriguing properties of TMDCs is their exceptionally large exciton binding energies. Due to the tight quantum confinement and reduced Coulomb screening, excitons in TMDCs exhibit binding energies up to several hundreds of meV, which essentially govern both optical and electric properties of TMDCs[1]. Moreover, due to the existence of two degenerate K and K' valleys, TMDCs are well known to possess a series of exciton species with complicated interactions[2–6]. The extremely high surface-to-volume ratio, on the other hand, makes the optical properties of TMDCs very sensitive to their surrounding environment[7–11]. As a result, TMDCs have become an ideal platform for the study of complex excitonic phenomena, such as excitonic insulators and Moire excitons.

Currently, studies on TMDCs are mostly focused on the relaxation dynamics of excitons, such as the formation of excitons, intervalley scattering, and exciton-phonon coupling[12–16]. However, it's worth pointing out that the transport / diffusion of excitons directly determines the performance limits of state-of-the-art excitonic devices[17]. Especially for monolayer TMDCs with strong excitonic effects, exciton transport may exhibit characteristics that are entirely different from those of traditional bulk materials. Indeed, a variety of fascinating exciton transport phenomena in TMDCs have been reported, including the exciton funneling effect[18,19], defect-induced hopping transport[20], effective negative diffusion, and transient superdiffusion[21–25]. Although significant progresses have been made in the quantitative characterization of the effective exciton diffusion coefficient, the microscopic mechanisms underlying the nonlinear diffusion behaviors in TMDCs remain elusive so far.

In particular, it was reported in recent years that monolayer $WS_2$ exhibits Halo patterns in its excitonic emission profiles under high pumping density[26–30]. To explain this phenomenon, it was initially proposed that some kind of memory effect, whose mechanism remains unclear to date, exists in the excitonic emission process of $WS_2$[27]. On the other hand, two other theoretical models which involve exciton drag by nonequilibrium phonons and Auger recombination induced excitonic temperature gradients, respectively, were introduced later[31,32]. However, the lack of further experimental support has kept these two models subjects of controversy. As a paradigm showing the highly nonlinear diffusion of excitons in TMDCs, clarifying the origin of such Halo patterns is of crucial importance for the fundamental understanding of exciton transport and its control in layered semiconductors.

In this work, we address this issue through high-precision time- and spatially-resolved photoluminescence imaging spectroscopy. In addition to the ultrafast spatial hole burning effect akin to the reported Halo patterns, we also observed the re-brightening of the emission hole under intense optical pumping. By combining time-resolved spectroscopy with ultrafast fluorescence imaging, we revealed that the emergence of spatial hole burning is invariably accompanied by an increase in the sample's doping level. Correspondingly, re-brightening of the emission hole is accompanied by significant de-doping of the sample. Building on these observations, we proposed a local doping model based on the ultrafast exciton-exciton annihilation (EEA) effect and the efficient capturing of EEA-generated holes by intrinsic sulfur vacancy defects in TMDCs. Excellent agreement between our theoretical calculations and the experimental data are found, thereby substantiating the validity of our model.

## Results

### Optical characterization of exciton states in $WS_2$

The samples we used are monolayer $WS_2$ flakes mechanically exfoliated from $WS_2$ single crystals (purchased from HQ, the Netherlands). The pumping source is a 515 nm femtosecond laser with a pulse width of ~150 fs and repetition rate of 76 MHz. The inset in Fig. 1(a) shows an optical image of a typical $WS_2$ flake. A smooth surface and regular shape are clearly visible, indicating the high quality of our samples. The thickness of the sample is scanned using atomic force microscopy (AFM), as demonstrated in Fig. 1(a). The extracted thickness is ~0.65 nm, confirming the monolayer nature of the flakes. A typical photoluminescence (PL) spectrum is plotted in Fig. 1(c). As one can see, the spectrum is composed of two components, with the dominant peak coming from neutral excitons ($X^0$) and the minor peak from charged excitons ($X^-$). To resolve the ultrafast diffusion dynamics of excitons, we carried out time-resolved PL imaging measurements using a streak camera (C16910, Hamamatsu). Typical results under average pumping power of ~1.2 μW are shown in the left panel of Fig. 1(b). From such a time-resolved image, we could obtain the evolution of the emission intensity, as demonstrated in Fig. 1(d). Through single exponential fitting, a PL lifetime of ~86.3 ps can be extracted, which is in general agreement with literatures. To demonstrate the diffusion of excitons more clearly, we normalized the time-resolved PL image to the intensity maximum at each time step, as shown in the right panel of Fig. 1(b). Obvious broadening of the emission spot can be observed, indicating the spatial diffusion of the optically injected carriers.

## Observation of ultrafast spatial hole burning and the re-brightening effect

To systematically study the exciton diffusion dynamics, we performed power-dependent and time-resolved PL imaging measurements. Typical results are shown in Fig. 2(a). For the lowest pumping power of ~0.52 μW, what we could observe is again the linear thermal diffusion of excitons, as expected. However, when pumping power was increased to ~2.08 μW, significant changes were observed. Emission from the center of the pumping spot was quenched ~100 ps after femtosecond laser pumping, generating a spatial hole burning pattern. As pumping power was further increased, the spatial hole burning area became larger and larger. On the other hand, the spatial hole burning pattern also appeared earlier and earlier with the increase of pumping power. Even more surprisingly, when pumping power reached ~607 μW, a sharp emission belt appeared in the center of the pumping spot that is originally dark. For high-enough pumping power of ~1012 μW, intensity of this emission belt even became much higher than those of the two strips lying in the outer region of the pumping spot. To further characterize the spatial hole burning dynamics and the re-brightening phenomena, we extracted the temporal evolution trace of the emission intensity from the time-resolved images, as shown in the upper panel of Fig. 2(b). As one can see, the luminescence lifetime decreases significantly with increasing pumping power. Furthermore, while the decay curve at low pumping power shows single-exponential behaviors, bi-exponential characteristics can be identified unambiguously at high pumping powers. The two decay curves in the lower panel of Fig. 2(b) show the results for pumping powers of ~607 μW and ~1012 μW, respectively. As pointed out, these two pumping powers are high enough to induce re-brightening of the emission hole. Interestingly, the decay curves show that now the luminescence lifetime turns to increase with the pumping power, in sharp contrast with the behavior at relatively low pumping power. The inset in the lower panel of Fig. 2(b) shows the power dependence of the two luminescence lifetimes extracted through bi-exponential fitting. As demonstrated, the luminescence lifetime decreases sharply as the spatial hole burning effect emerges and turns to increase with the pumping power after the re-brightening process of the hole starts.

Multi-dimensional spectroscopic analyses

Having revealed the key characteristics of the ultrafast spatial hole burning dynamics, we now turn to its microscopic mechanisms. To do this, we carried out spatially-, energy-, and time-resolved multi-dimensional spectroscopic measurements. As shown in Fig. 3(a), we selected three key positions for further energy- and time-resolved spectroscopic studies. Position B is located at the center of the pumping spot. Positions A and C correspond to the two outer bright strips. The three figures on the right of Fig. 3(a) show the time-resolved PL images for A, B and C. It can be identified unambiguously that Position B has a very short PL lifetime, while lifetimes for A and C are rather long. Fig. 3(b) shows the corresponding energy- and time-resolved PL images for these three selected positions. As one can see, emissions from the two outer strips are dominated by a high-energy peak at ~612 nm, while emission from the pumping spot center is dominated by a relatively low-energy peak at ~620 nm. Such spectral differences can be demonstrated even more clearly from their PL spectra, as plotted in Fig. 3(c). Together shown in Fig. 3(c) are the double-peak fittings using two Gaussian functions. According to the characteristics of the fitted peaks and existing literature, we could identify that the high-energy peak at ~612 nm comes from neutral excitons, and the low-energy peak at ~620 nm originates from negatively-charged excitons. These results clearly indicate that the center of the pumping spot has much higher doping level than its surrounding area. As trions decay predominantly via non-radiative channels[33], emission intensity from the pumping spot center is much lower than the surrounding area, thus generating spatial hole burning patterns.

Based on the above discussion, we propose a local doping model for the observed ultrafast spatial hole burning effect. As shown schematically in Fig. 3(f), under intense optical pumping, exciton-exciton annihilation (EEA), which can be regarded as the excitonic version of Auger recombination, becomes significant[34–36]. A considerable number of free electrons and holes are thus generated. However, due to the existence of sulfur vacancy defects which are well known to serve as centers of negative charges[37–39], holes generated from the EEA process are readily attracted and captured by these sulfur vacancies, leading to the increase of doping concentration in the sample[40,41]. As the center of the pumping spot has the most pronounced EEA effect and thus the highest doping level, excitons in the spot center are efficiently transferred into trions which decay primarily non-radiatively, leading to spatial hole burning patterns.

Here, we also notice that the originally dark pumping spot center can be re-brightened under sufficiently high pumping power. To reveal the underlying physics, we again performed spatially-, energy- and time-resolved PL imaging analyses. Typical results are plotted in Fig. 3(d). The left panel in Fig. 3(d) shows the spatially- and time-resolved image under high pumping power of ~1.82 mW, where re-brightening of the spot center can be clearly identified. The corresponding energy- and time-resolved image is given in the right panel of Fig. 3(d). To clearly demonstrate the spectral changes after the spot center is re-brightened, the above-threshold PL spectrum is plotted together with that taken below threshold, as shown in Fig. 3(e). As one can see, the above-threshold spectrum is dominated by neutral exciton emission while the below-threshold spectrum is dominated by trions. These observations indicate that the doping concentration decreases significantly after the spot center is re-brightened. Referring to the intrinsic properties of TMDCs and literature, we suggest that such a de-doping effect originates from the photo-oxidation of the sample under intense laser pumping. With the assistance of intense laser irradiation, some of the sulfur atoms in WS$_2$ can be replaced by the oxygen atoms from O$_2$ and H$_2$O molecules[42–44]. This causes significant de-doping of the monolayer flakes, thus re-brightening the spot center. A schematic diagram showing such photo-oxidation (chemical adsorption) as well as physical adsorption and desorption is given in Fig. 3(g).

**Discussion**

To further support our model, we carried out theoretical simulations by taking into account the EEA process, capturing of holes by intrinsic sulfur vacancy defects, and thermal diffusion of excitons. For simplicity, we consider a one-dimensional model. In this case, equations describing our model can be written as follows:

$$\frac{\partial n_x(x,t)}{\partial t} = D\Delta_x n_x - \frac{n_x}{\tau_x} - R_A n_x^2 - T n_x n_e \tag{1}$$

$$\frac{\partial n_e(x,t)}{\partial t} = R_A n_x^2 + D_e \Delta n_e - \frac{n_e}{\tau_e} \tag{2}$$

Here, $n_x(x, t)$ describes the distribution of neutral excitons in time- and spatial domain, $D$ is the effective diffusion coefficient of neutral excitons, $\Delta_x$ is the Laplace operator, $\tau_x$ represents the lifetime of neutral excitons, $R_A$ describes the EEA rate, and $T$ is the rate of neutral excitons being converted into trions. As the intensity of the pumping spot has a Gaussian profile, the initial density of neutral excitons is described by a Gaussian function $n_x(x,0) = P * e^{-\mu*(x-0)^2}$. Equation (2) describes the dynamics of free electrons generated by laser pumping, with $n_e$ being the density of free electrons. $D_e$ is the diffusion coefficient of free electrons. Here, it should be noted that the holes captured by sulfur vacancy defects will eventually be released and subsequently recombine with free electrons. The last term on the right-hand side of Equation (2) describes this physical process, with $\tau_e$ being the statistical lifetime of photo-generated electrons. Moreover, we also note that photo-oxidation occurs only when the pumping density is higher than the threshold. As an approximation, we could quantify the extent of oxidation of the sample by the difference between the local exciton concentration and a threshold concentration, $\Delta n(x,t) = n_x(x,t) - n_{th}$. As photo-oxidation de-dopes the monolayer flakes, the density of free electrons in the case of photo-oxidation can be expressed as $n_e(x,t) - \xi \cdot (n_x(x,t) - n_{th})$, with $\xi$ being the de-doping coefficient.

Typical simulation results are plotted in Fig. 4(a). As one can see, our model can reproduce both the ultrafast spatial hole burning and the re-brightening effects very well. Quantitative comparison between theoretical calculations and our experimental data is shown in Fig. 4(b). The left panel in Fig. 4(b) shows the spatial profile of the PL signal at different moments, under an average pumping power of ~205 µW. Ultrafast spatial hole burning of the excitonic emissions can be clearly identified. In the right panel of Fig. 4(b), we show the PL profiles taken at 113 ps under different pumping powers. Spatial hole burning in the spot center and its re-brightening can be clearly identified. Most importantly, our calculations show excellent agreement with the experimental data for both spatial hole burning behavior and the re-brightening effect. Time-dependent densities of neutral excitons and free electrons are plotted in Fig. 4(c) for below-threshold pumping (without photo-oxidation) and in Fig. 4(d) for above-threshold pumping (with photo-oxidation), respectively. For below-threshold pumping, the fast-increasing concentration of free electrons at the initial stage, which is due to the pronounced EEA effect, accounts for the formation of spatial hole burning patterns. For above-threshold pumping, the density of free electrons decreases quickly at the initial stage, manifesting the strong photo-oxidation effect under intense femtosecond laser pumping. When the exciton concentration falls below the threshold, the photo-oxidation process ceases, and the concentration of free electrons begins to increase over time. Here, it is noteworthy that, compared with the spatial hole burning stage, the overall level of free electron concentration during the photo-oxidation stage is significantly lower. This explains why neutral excitons dominate the PL spectra during the photo-oxidation stage. Figs. 4(e) and 4(f) show typical calculated spatial distribution of neutral excitons and photo-generated electrons at different moments under below-threshold and above-threshold pumping, respectively. As one can see, the anomalous exciton diffusion is obviously correlated with the spatial distribution of photo-generated electrons during both the formation of the hole burning effect and the oxidation at the center of the pumping spot.

Finally, we would like to further discuss the reversibility of the observed spatial hole burning effect and the photo-oxidation process. For the ultrafast spatial hole burning effect, one may argue that it could result from degradation of sample quality at the center of the pumping spot due to the high pumping density. This possibility can be excluded based on our reversibility measurements. As demonstrated in Fig. S2 (supplementary information), when we reduced the pumping power the emission pattern fully returns to its initial stage before the emergence of the spatial hole burning effect. However, at sufficiently high pumping power where photo-oxidation is triggered, the emission pattern is found to be irreversible, as shown in supplementary Fig. S3. This is consistent with our proposed model. As substitution of lattice atoms is involved, the photo-oxidation process is itself predicted to be irreversible.

In summary, we report systematic studies on the ultrafast spatial hole burning effect in monolayer $WS_2$. We revealed that the pronounced exciton-exciton annihilation process, together with the efficient capturing of holes by intrinsic sulfur vacancy defects, induces significant photo-doping in the pumping spot center and causes ultrafast spatial hole burning in excitonic emission. At sufficiently high pumping power, we observed re-brightening of the spot center and confirmed that it originates from the laser induced photo-oxidation. Excellent agreements are found between our diffusion model calculation and the experimental results. These results reveal the microscopic mechanism for the nonlinear ultrafast diffusion of excitons and its optical control in monolayer TMDCs, opening new ways for the development of layered semiconductor-based excitonic devices.

**Methods**

Sample preparation

Bulk $WS_2$ crystals were mechanically exfoliated using a low-adhesion blue tape. To obtain high-quality monolayer flakes, the exfoliated fragments were repeatedly laminated and peeled using several pieces of clean, weakly adhesive blue tape, which were then pressed onto a polydimethylsiloxane (PDMS) substrate. After resting for ~5 minutes, the tape was swiftly removed from the PDMS surface, leaving behind monolayer $WS_2$ flakes with lateral dimensions exceeding 100 μm. These monolayers were subsequently transferred onto clean $SiO_2$ substrates using a two-dimensional dry transfer stage for further characterization.

Optical characterization system

Photoluminescence spectra were acquired using an iHR550 spectrometer (Horiba Jobin-Yvon) equipped with a 150 grooves/mm grating blazed at 500 nm, yielding a spectral resolution of ~0.47 nm. Time-, energy-, and spatially-resolved photoluminescence imaging measurements were performed using a home-built confocal micro-spectroscopic system equipped with a streak camera (C16910, Hamamatsu). The system offers a spectral response range from 350 to 850 nm, a spatial resolution of approximately 15 lp/mm, and a temporal resolution of ~2 ps.

## Data Availability

The data that support the findings of this study are available from the corresponding authors on reasonable request.

## Acknowledgements

This work was financially supported by the National Natural Science Foundation of China (Grant No. 12274159), and the National Key Research and Development Program of China (Grant No. 2022YFA1602700).

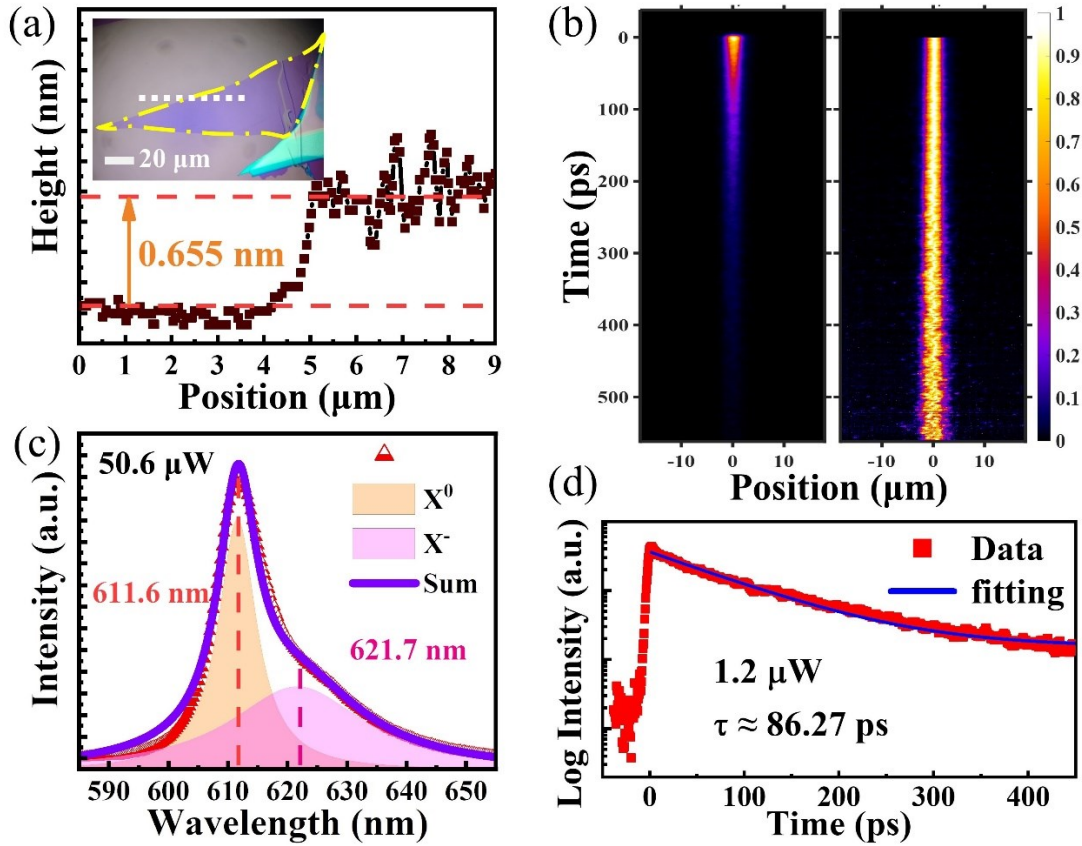

**FIG. 1. Sample and time-resolved characterization.** (a) Atomic force microscopy (AFM) measurements of a typical mechanically exfoliated WS$_2$ monolayer flake. Inset: optical image of the flake. The white dotted line in the inset marks the AFM scanning direction. (b) Left panel: typical time-resolved PL image pumped using a 515 nm femtosecond laser. Right panel: the same data as shown in the left panel, but the data have been normalized to the intensity maximum at each time step. (c) Typical PL spectrum measured under average pumping power of 50.6 μW and its double-peak fitting. (d) Evolution of the PL intensity in the time domain and its single-exponential fitting.

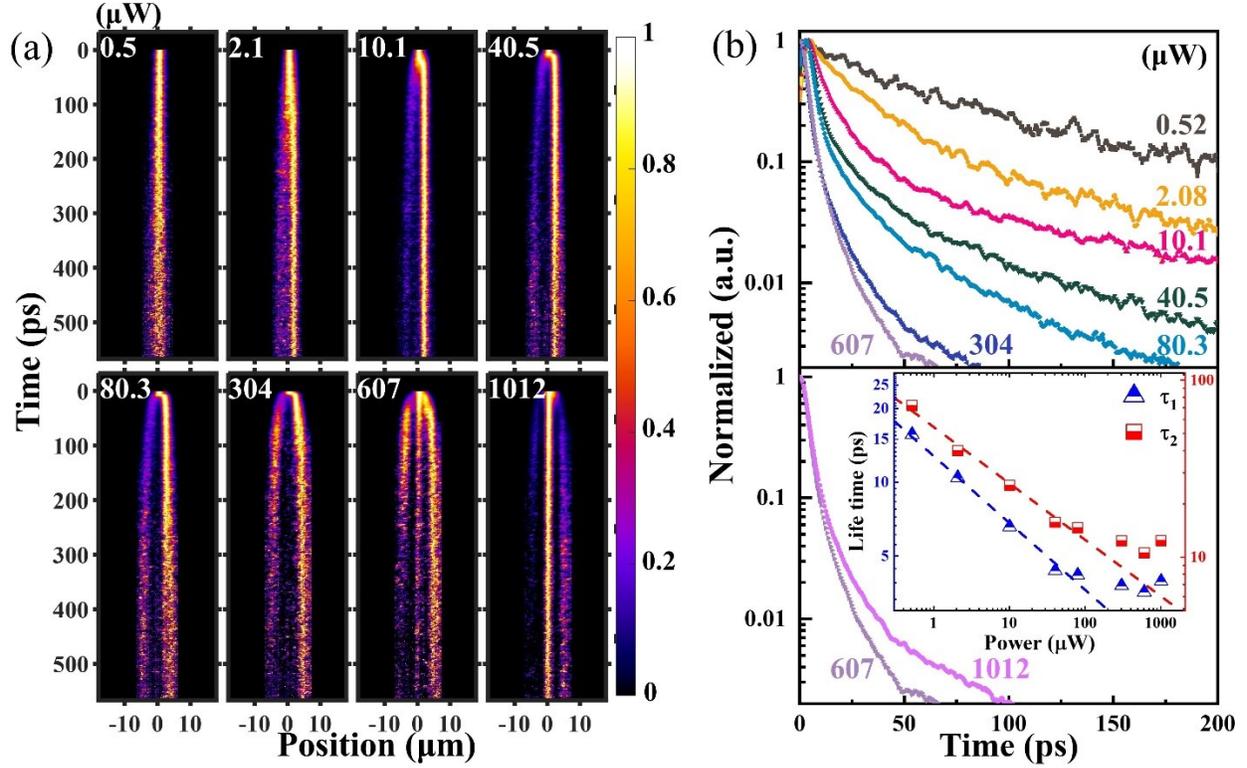

**FIG. 2. Ultrafast spatial hole burning and the re-brightening effect.** (a) Time-resolved PL images of the monolayer WS$_2$ flake under different pumping powers. The numbers in the images indicate the average pumping power of the 515 nm femtosecond laser (in unit of μW). (b) Upper panel: evolution of the PL intensity in the time domain under different pumping powers ranging from 0.52 μW to 607 μW. Lower panel: comparison of the PL intensity evolution curves under pumping power of 607 μW and 1012 μW. Inset: power dependence of the PL lifetimes $\tau_1$ and $\tau_2$ extracted through bi-exponential fitting. Note that both the horizontal and vertical axes use logarithmic scales. The two dashed lines in the inset are guidelines to highlight the increase of the PL lifetimes at sufficiently high pumping power.

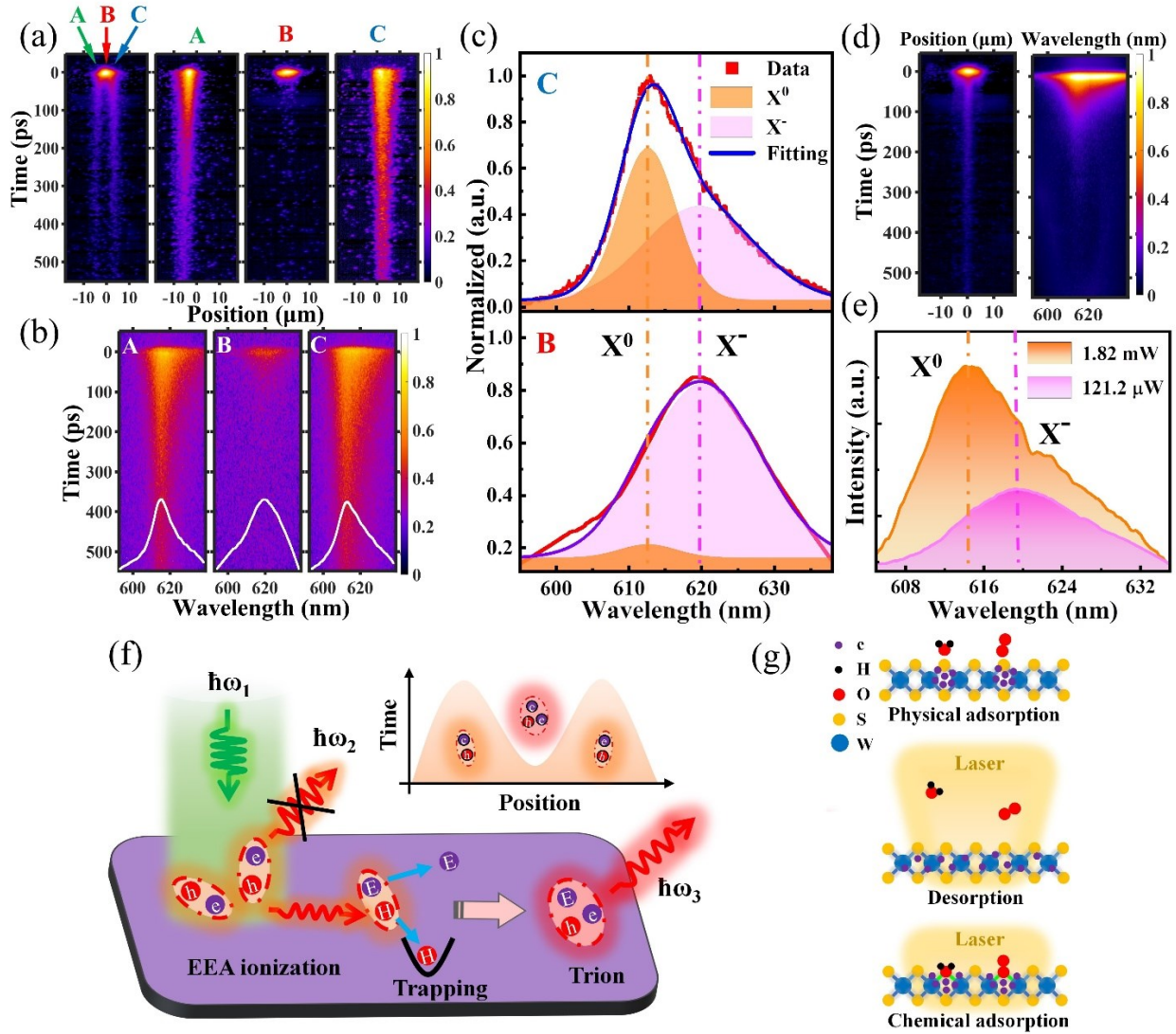

**FIG. 3. Multi-dimensional spectroscopic analyses.** (a) Time- and spatially-resolved PL images of $WS_2$. A, B, and C mark the three key positions selected for further energy- and time-resolved imaging analyses. B corresponds to the center of the pumping spot. A and C correspond to the two bright outer strips. (b) Energy- and time-resolved PL images for the three selected detection positions A, B, and C, respectively. The white solid curves superimposed on the images are the corresponding PL spectra obtained by integrating the time-resolved images. (c) Upper panel: PL spectrum for position C and its double-peak fitting. Lower panel: PL spectrum and its double-peak fitting for position B. (d) Left (right) panel: spatially- (energy-) and time-resolved PL images at sufficiently high pumping power where re-brightening of the spot center happens. (e) PL spectra for below threshold (121.2 μW, without re-brightening effect) and above threshold (1.82 mW, with re-brightening effect) pumping, respectively. (f) Schematic diagram showing the mechanism of local doping induced by laser pumping. (g) Schematic diagram showing the three different types of interactions between gas molecules and TMDCs, including physical adsorption, desorption, and chemical adsorption.

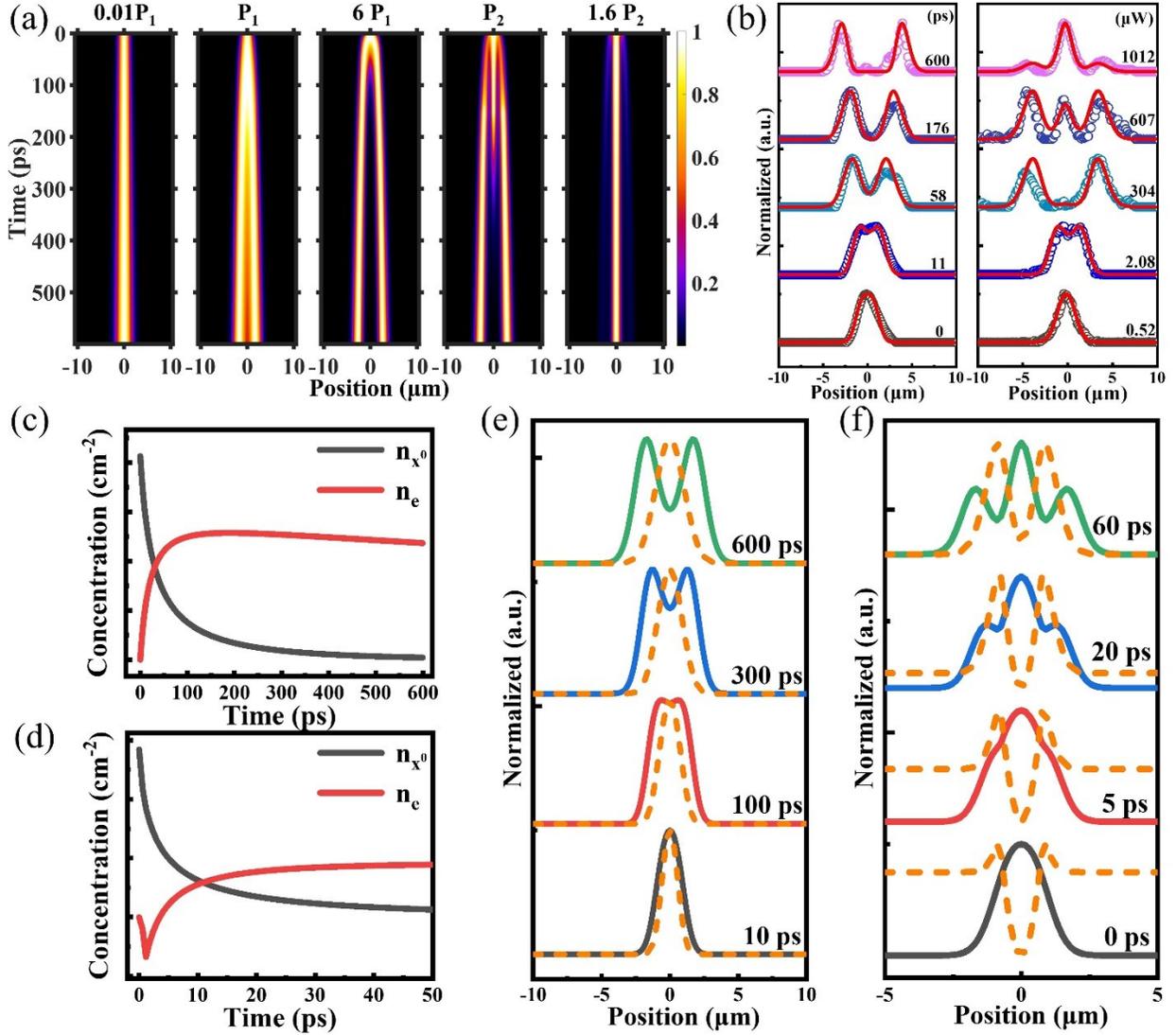

**FIG. 4. Theoretical simulations.** (a) Simulated time- and spatially-resolved PL images under different pumping powers. $P_1$ denotes the threshold power under which the spatial hole burning effect can be triggered. (b) Comparison between theoretical calculations and experimental data. Left panel: spatial profiles of the PL signal taken at different moments. Right panel: spatial profiles of the PL signal taken at 113 ps but under different pumping powers. Circles: experimental data. Solid curve: theoretical calculations. (c) Calculated densities of neutral excitons and photo-generated electrons as a function of time under pumping power where spatial hole burning effect is triggered. (d) Calculated densities of neutral excitons and photo-generated electrons under sufficiently high pumping power where re-brightening of the spot center happens. (e) Calculated spatial distribution of neutral excitons and photo-generated electrons at different moments under pumping power where spatial hole burning effect is triggered. (f) Calculated spatial distribution of neutral excitons and photo-generated electrons at different moments under sufficiently high pumping power where re-brightening of the spot center happens.

# Ultrafast Spatial Hole-burning Dynamics in Monolayer WS$_2$: Insights from Time-resolved Photoluminescence Spectroscopy

## Supplementary Information

Sample preparation and measurement system

The samples we used are monolayer WS$_2$ flakes mechanically exfoliated from WS$_2$ single crystals (purchased from HQ, the Netherlands). Time-, energy-, and spatially-resolved PL imaging measurements were carried out using a home-built confocal micro-spectroscopic system equipped with a streak camera (C16910, Hamamatsu). A schematic diagram of this imaging system is shown in Fig. S1. The pumping source is a 515 nm femtosecond laser (Pharos, Light Conversion) with pulse width < 290 fs, repetition rate of 76 MHz and maximum single pulse energy of 200 μJ. A 50X objective (Olympus) with NA of 0.5 is used to image the samples and focus the laser beam onto sample surface (spot diameter: ~3 μm). A confocal hole with diameter of 50 μm is used to increase the spatial resolution of the system. A 561 nm long-pass filter (Semrock) is used to filter out relected laser light. All measurements were carried out at room temperature.

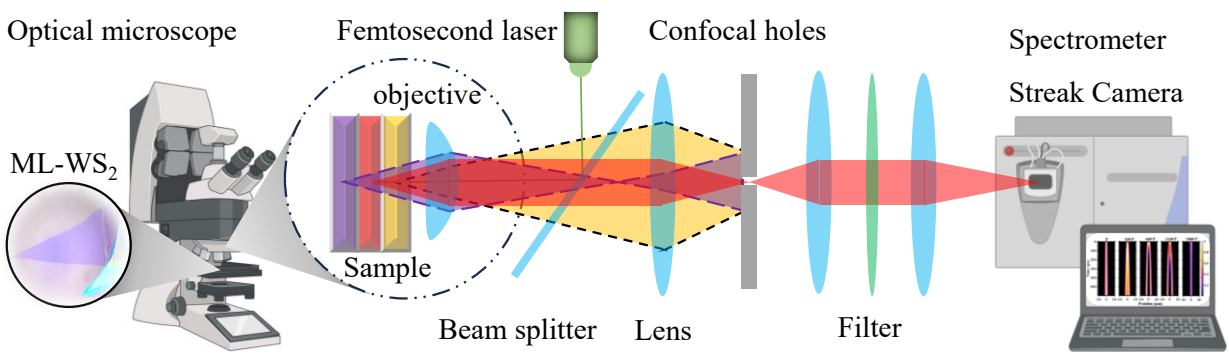

**Fig. S1 | Confocal Micro-spectroscopic Measurement System.** The optical microscope and spectrometer icons in the figure were adapted from BioRender.com.

# Reversibility of the ultrafast spatial hole burning effects and photo-oxidation processes

Fig. S2(a) shows the time- and spatially-resolved PL images under different pumping powers. We firstly increase the pumping power from 0.48 µW to 80.6 µW, and then reduce it down to 0.52 µW. As one can see, the spatial hole burning effect becomes more and more significant with the pumping power in the first stage. However, as we reduce the pumping power, the spatial hole burning pattern disappears again. Fig. S2(b) shows the evolution of the PL intensity in the time domain when the pumping power was increased gradually. Obviously, the curve changes from single-exponential decay at low pumping power to bi-exponential decay at high pumping power. Fig. S2(c) shows the corresponding data in the down-sweep measurement. Clearly, the curve changes back from bi-exponential decay at high pumping power to sing-exponential decay at low pumping power. These data confirm that the spatial hole burning effect observed in this work is fully reversible.

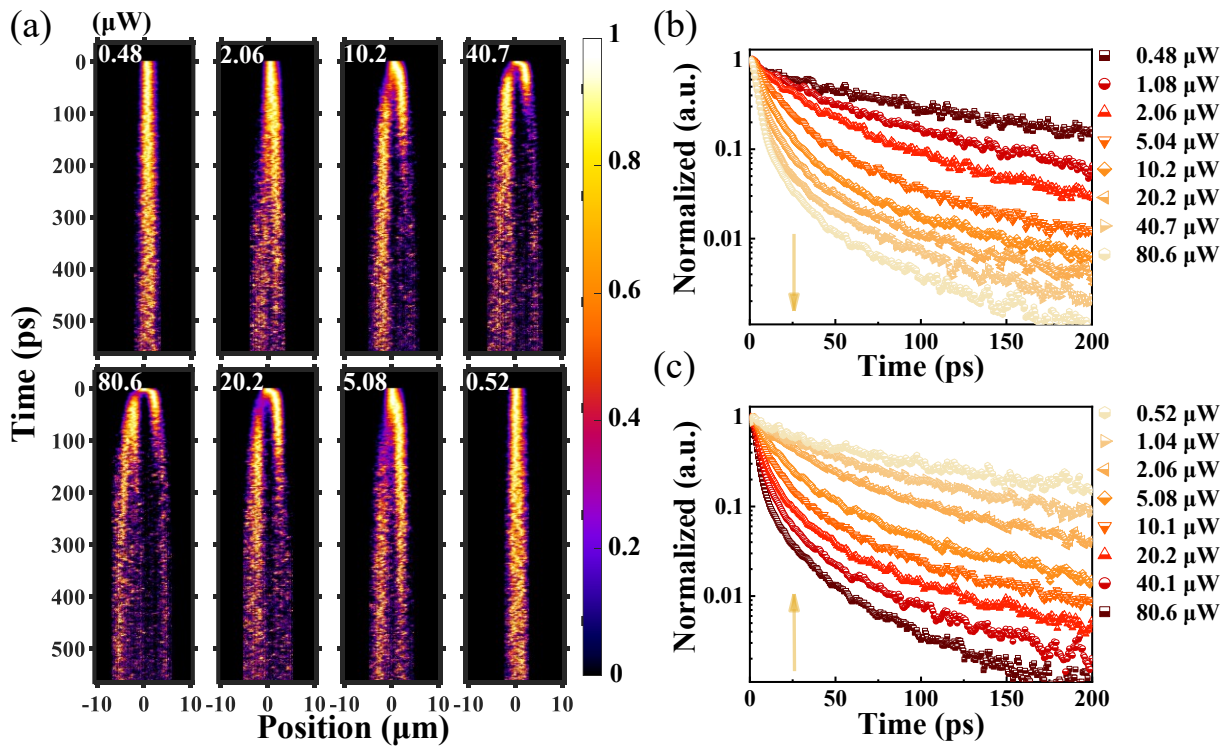

**Fig. S2 | Reversibility of the spatial hole-burning process. a** Time- and spatially-resolved PL images under different pumping power. The pumping power was firstly increased from 0.48 µW to 80.6 µW, and then reduced to 0.52 µW.

**b** Evolution of the PL intensity under different pumping powers in the up-sweep measurements. **c** Evolution of the PL intensity under different pumping powers in the down-sweep measurements.

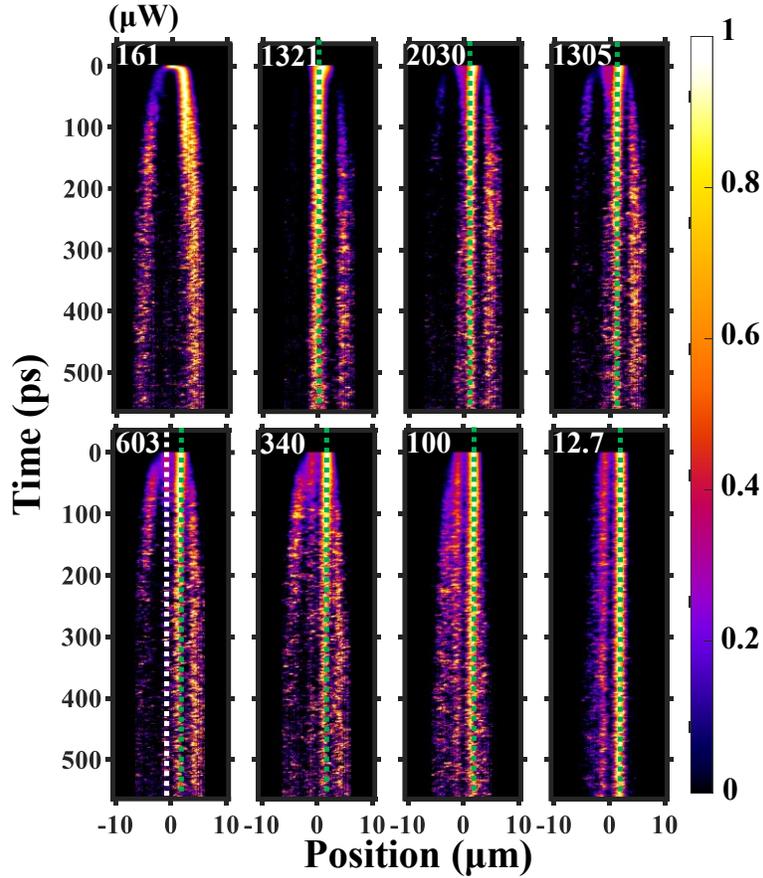

**Fig. S3 | Irreversible re-brightening effect in the spot center.** Time- and spatially-resolved PL images under different pumping powers. The pumping powers used here are much higher than those in Fig. S2. The power was firstly increased from 161 µW to 2030 µW and then reduced to 12.7 µW. The dotted lines highlight the re-brightening positions under sufficiently high pumping power.

Fig. S3 shows another set of time- and spatially-resolved PL images under different pumping powers. The pumping powers used here are much higher than those in Fig. S2. In addition to spatial hole burning, re-brightening effects in the spot center can also be clearly identified, as highlighted by the dotted lines. As one can see, even when the pumping power was reduced to 12.7 µW (which is much lower than the initial power of 161 µW), the emission patterns cannot be restored to their initial state.

# Time-space PL imaging under high vacuum

Fig. S4 shows the time- and spatially-resolved PL images taken under pressure of 1.5 Torr. As one can see, even the pumping power is much higher than those in Fig. S2, we didn't observe the spatial hole burning effect. This is because $H_2O$ and $O_2$ molecules, which are originally adsorbed onto $WS_2$ surface, desorb in high vacuum environment. Such desorption increases the doping level of the $WS_2$ flakes, making more neutral excitons transfer into trions. As trions decay primarily via non-radiative channels, the EEA effect is strongly surpressed. As a result, the spatial hole burning effect can no longer be observed. Another feature in Fig. S4 is that the spot center shows a weak but long strip under very high pumping power (1202 µW and 2010 µW). This strip results from the photo-oxidation process in the spot center. As the vacuum level used in this experiment is not very high (1.5 Torr), there are still some $H_2O$ and $O_2$ molecules adsorbing onto the sample surface. Under the illumination of high power laser, these adsorbed molecules can still cause photo-oxidation of the sample.

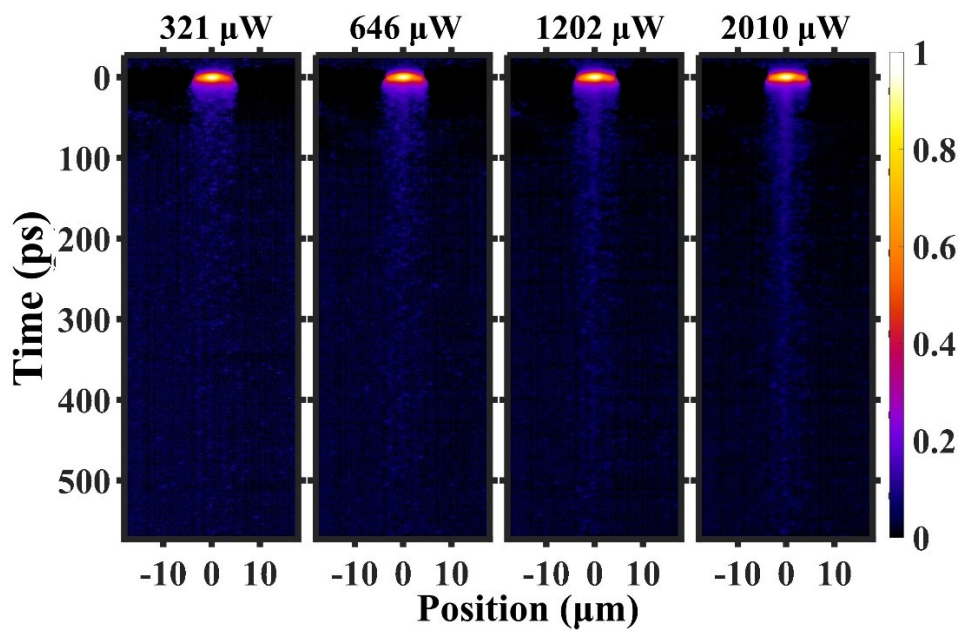

**Fig. S4 | Time-space PL imaging under high vacuum.** Time- and spatially-resolved PL images measured under pressure of 1.5 Torr. Pumping powers are listed on the top of each image.

# Reproducibility of the observations

The ultrafast spatial hole burning and the re-brightening effects are commonly observed in our samples. Fig. S5 shows the time- and spatially-resolved PL images in another monolayer WS$_2$ flake. Similar spatial hole burning and re-brightening phenomena can be clearly identified.

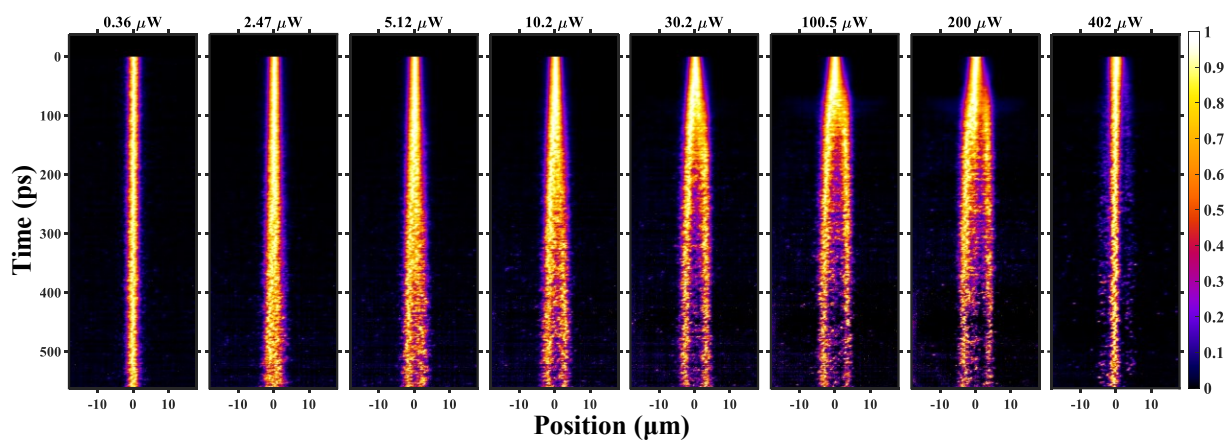

**Fig. S5 | Ultrafast spatial hole burning and the re-brightening effect in another WS$_2$ flake.**

Meanwhile, as monolayer semiconductors with extremely large surface-to-volume ratio, optical properties of monolayer TMDCs are very sensitive to their surrounding environment. To test the robustness of the ultrafast spatial hole burning effect, we carried out measurements on WS$_2$ flakes deposited on non-hydrophilic substrate of PDMS. Typical results are plotted in Fig. S6. As one can see, by increasing the pumping power, ultrafast spatial hole burning and the re-brightening effect are again observed. Compared with WS$_2$ flakes on SiO$_2$ substrate, the main difference is that the threshold power for spatial hole burning is significantly higher. To further study the effect of substrate on the nonlinear diffusion behaviors, we tried to extract the effective diffusion coefficients. Fig. S7(a) shows the spatial profiles of the PL signal at different moments for WS$_2$ flakes deposited on SiO$_2$ and PDMS substrates, respectively. Fig. S7(b) shows the square of the FWHM of the PL spatial profile as a function of time. Fitting the data using the function $\omega^2(t) = \omega^2(t) + 4D_{eff}t$, we could extract the effective diffusion coefficient, which is $D_{\text{eff}}$ = 19.15 cm$^2$/s for WS$_2$ on PDMS substrate and $D_{\text{eff}}$ = 47.3 cm$^2$/s for WS$_2$ on SiO$_2$, respectively. Owing to the low van der Waals binding energy between PDMS and monolayer WS$_2$, the contact between WS$_2$ and PDMS is weaker than that between WS$_2$ and the SiO$_2$ substrate. As a result, the exciton diffusion coefficient for SiO$_2$ substrate is significantly larger than that for PDMS substrate.

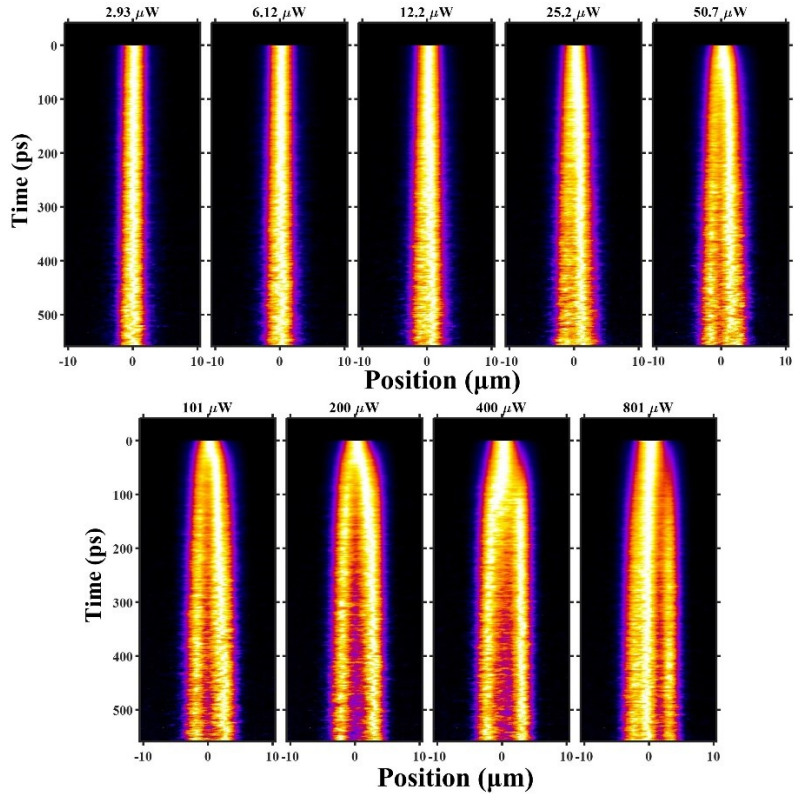

**Fig. S6 | Time- and spatially-resolved PL images of monolayer WS$_2$ on non-hydrophilic substrate of PDMS.** Ultrafast spatial hole burning and re-brightening effects can be clearly observed.

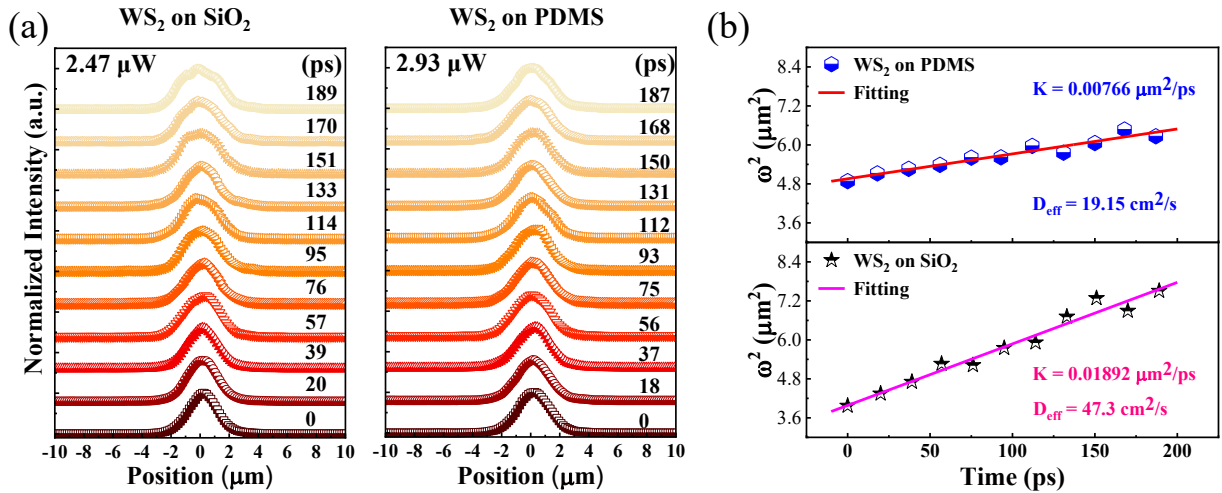

**Fig. S7 | Effect of substrates on exciton diffusion in monolayer WS$_2$. a** Spatial profiles of PL signal at different moments for WS$_2$ deposited on SiO$_2$ (left panel) and PDMS (right panel) substrates, respectively. **b** Square of the FWHM of the PL spatial profile as a function of time. The solid lines are theoretical fittings.

# Fitting parameters for our diffusion model

**Table 1 Fitting parameters for the ultrafast spatial hole burning dynamics**

| Symbols | Initial Values | Units | Comments |
|---|---|---|---|
| $n_x$ | $10 * e^{-0.91*(x-0)^2}$ | $\mu m^{-2}$ | Neutral exciton concentration |
| $\Delta_x = \frac{\partial}{\partial x^2}$ | — | — | 1D Laplace operator |
| $D$ | $5.3 \times 10^{-4}$ | $\mu m^2/ps$ | Neutral exciton diffusion rate |
| $\tau_x$ | 400 | $ps$ | Neutral exciton lifetime |
| $R_A$ | $0.5 \times 10^{-4}$ | $\mu m^2/ps$ | Neutral exciton EEA rate |
| $T$ | $1 \times 10^{-5}$ | $\mu m^2/ps$ | Neutral exciton to trion conversion rate |
| $n_e$ | 0 | $\mu m^{-2}$ | Concentration of excess electrons generated by photo-doping |
| $\tau_e$ | 4000 | $ps$ | Lifetime of excess electrons generated by photo-doping |
| $D_e$ | $5.3 \times 10^{-4}$ | $\mu m^2/ps$ | Diffusion rate of excess electrons generated by photo-doping |
| $n_{th}$ | 6000 | $\mu m^{-2}$ | Threshold concentration for photo-oxidation |
| $\xi$ | 1.2 | $\mu m^2/ps$ | De-doping coefficient due to photo-oxidation |